\documentclass[aps,prb,showpacs,twocolumn,amsmath]{revtex4}
\usepackage{graphicx}
\pdfoutput=1
\newcommand{\bq}{\begin{equation}}
\newcommand{\eq}{\end{equation}}
\newcommand{\bn}{\begin{eqnarray}}
\newcommand{\en}{\end{eqnarray}}
\newcommand{\bsub}{\begin{subequations}}
\newcommand{\esub}{\end{subequations}}

\newcommand{\al}{\alpha}
\newcommand{\eps}{\epsilon}

\begin{document}
\title{How to Measure the Transmission Phase via a Quantum Dot in a Two-Terminal Interferometer}
\author{Vadim I. Puller$^1$}
\author{Yigal Meir$^{1,2}$}
\affiliation{$^1$Department of Physics, Ben-Gurion University of
the Negev, Beer Sheva 84105 Israel\\$^{2}$ The Ilse Katz Center
for Meso- and Nano-scale Science and Technology, Ben-Gurion
University, Beer Sheva 84105, Israel}

\begin{abstract}
{Measurement of the transmission phase through a quantum dot (QD) embedded in
an arm of a two-terminal
Aharonov-Bohm (AB) interferometer is inhibited by phase symmetry, i.e.
the property that the linear response conductance of a two-terminal device is
an even function of magnetic field. It is  demonstrated that  in a setup consisting of an
interferometer with a QD in each of its arms, with one of the QDs
capacitively coupled to a nearby quantum point contact (QPC),
phase symmetry is broken when a finite voltage bias is applied to the QPC. The  transmission phase via the uncoupled QD  can then be deduced from the amplitude of the odd component of the AB oscillations.}

\end{abstract}
\date{\today}

 \maketitle

Measuring the transport amplitude $t$ through a quantum dot (QD), as a function of energy (or gate voltage), can give detailed information about the energy structure, the wave functions and the many-body correlations in the QD. While the absolute value of the transmission $|t|^2$ can be easily measured,\cite{QDreview} and has been employed extensively to characterize the QD, measurements of the transmission phase are much more subtle. The standard experimental approach is to embed the QD in an Aharonov-Bohm (AB) interferometer,\cite{Yacoby,YacobyPR,Schuster,Holleitner,JI,Kobayashi,Sigrist,Kalish,Leturcq,Zaffalon,Litvin} and study the change in the the phase of the AB oscillations, as a function of the QD parameters, e.g. gate voltage. However, the relation between the AB phase and the transmission phase is not straightforward. \cite{Amnon_open} In particular, when the AB interferometer is connected to two terminals, then the Onsager-B\"{u}ttiker relations dictate that the linear response conductance must be an even function of magnetic flux, \cite{Onsager,Buttiker,LeviYeyati} which means that the phase of AB oscillations can assume only the values $0$ or $\pi$ ({\em phase symmetry}), independent of the transmission phase through the QD.
This obstacle has been overcome experimentally by employing an
open interferometer, i.e. an interferometer with more than two leads. However, this approach requires advanced technology and  suffers from low signal due to particle losses to the other leads, and so far only a single group has been reporting measurements of the transmission phase, using this approach.\cite{Schuster,JI,Kalish,Zaffalon} Alternatively, it has been suggested that the phase may be extracted from a multi-parameter fit to the shape of AB oscillations \cite{Amnon_fit}.

\begin{figure}[tbp]
  \includegraphics[width=2in]{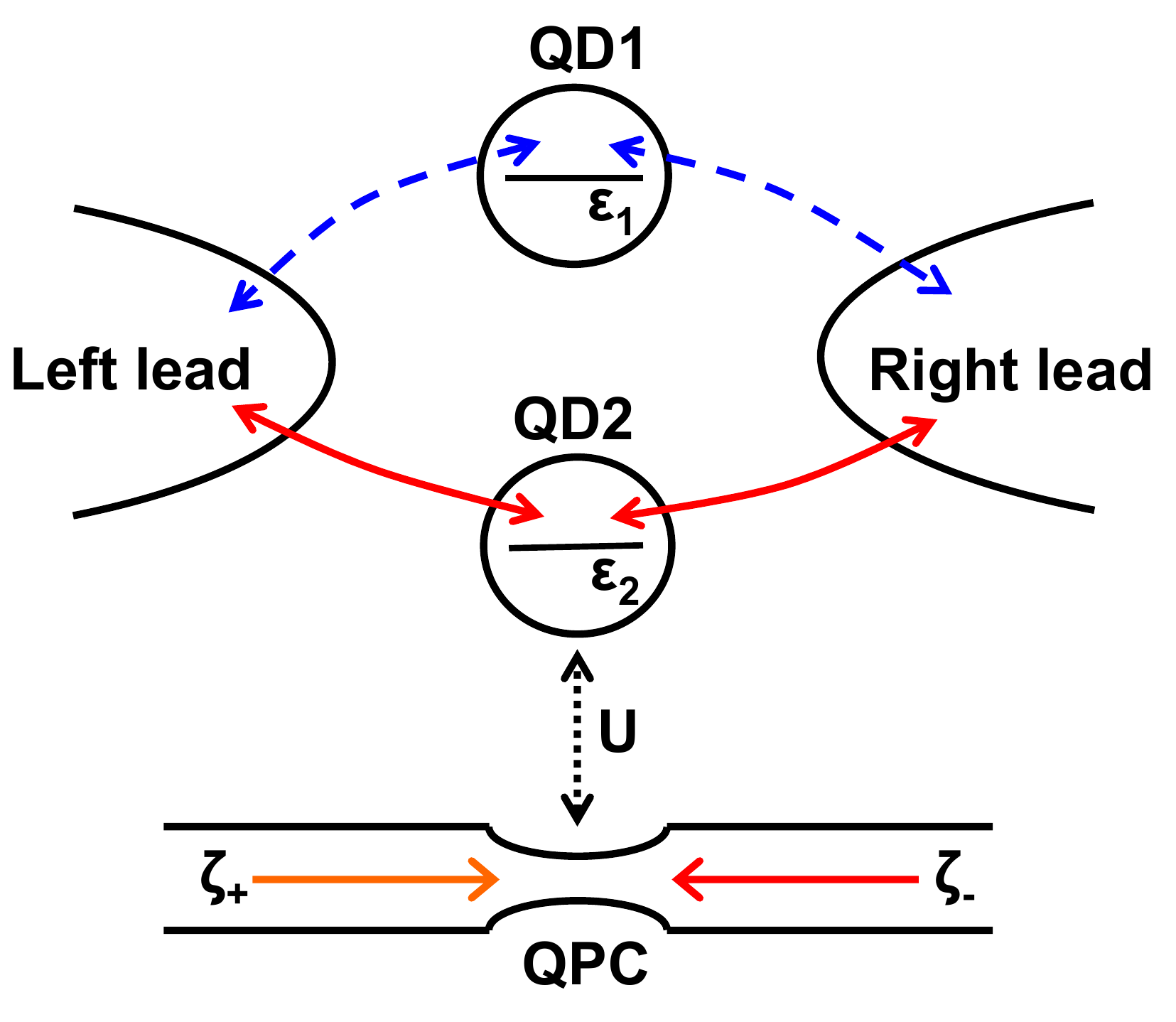}\\
  \caption{(color online) Schematics of "Which path?"
  interferometer studied in this paper. The Aharonov-Bohm interferometer contains a quantum dot in each arm. In order to measure the transmission phase through one of the dots (QD1), a quantum point contact is coupled electrostatically to the other dot (QD2), in order to break the phase symmetry.}\label{fig:setup_phase}
\end{figure}

In this Letter we propose a new way to measure directly the transmission phase via a QD in a two-terminal AB interferometer, by coupling it to a near-by quantum point contact (QPC). The proposed measurement setup (Fig.~\ref{fig:setup_phase}) resembles the so-called "Which path?" interferometer, \cite{Buks,Aleiner} which in our case consists of a two-terminal AB interferometer containing a QD in each of its arms, and a QPC capacitively coupled to one of the QDs (QD2), as shown in Fig.~\ref{fig:setup_phase}. The QPC is expected to reduce the amplitude of the AB oscillations,\cite{Aleiner} but more importantly in the present context, it causes breaking of the phase symmetry when a finite bias is applied to it. (This, in fact, is a special case of breaking of the phase symmetry due to a nonequilibrium environment.\cite{SanchezKang}) This phase-symmetry  breaking will enable a direct measurement of the transmission phase through the QD not coupled to the QPC (QD1).

In this geometry, and under QPC bias, when a level in QD1 is swept across the Fermi level of the interferometer leads, the AB phase smoothly flows between the values $0$ and $\pi$. Thus, breaking of the phase symmetry by coupling to a non-equilibrium environment can be observed experimentally, which, to our knowledge, has not been done so far.
The observed phase of the AB oscillations, however, is not the transmission phase via  QD1. Nevertheless, as we demonstrate below, the transmission phase through the QD1 can  still be extracted from the amplitude of the odd component of the AB oscillations.
In the following we  present the model describing
our "Which path?" detector, discuss breaking of the phase symmetry,  and propose a straightforward method of measuring the transmission phase via the QD.

We describe the system by the following Hamiltonian
\bn
{\cal H} &=& \sum_{\al}\eps_\al d_{\al}^\dagger d_\al +\sum_{k\in L,R}\eps_{k}c_{k}^\dagger c_{k}+\sum_{p,\nu}E_{p\nu}a_{p\nu}^\dagger a_{p\nu}\\
&+&\sum_{\al,k,\mu}\left(t_{\al\mu}d_\al^\dagger c_{k\mu}+h.c.\right)
+Ud_2^+ d_2\sum_{p,\nu;p'\nu'}a_{p\nu}^\dagger a_{p'\nu'}.\nonumber
\label{eq:Hamiltonian}
\en
The first term describes the non-interacting QDs forming the interferometer -- $d_{\al}$ is the operator that destroys an electron in QD $\al (\al=1,2)$. For simplicity we treat a single level in each dot, but the calculation and the results can be trivially extended to multi-level dots. The second term describes the leads: $c_{k}$ destroys an electron in state $k$. The states  in the leads are filled up to their respective chemical potentials $\mu_{L,R}$. The third term describes the QPC, the states in which are taken to be right/left-movers ($\nu=\pm$) labeled by wave numbers $p,p'$. $a_{p\nu}$ is the corresponding electron destruction operator. Right- and left-moving bands are filled up to their respective chemical potentials $\zeta_{\nu}$.
The fourth term accounts for tunneling between the leads and the QDs, characterized by coupling strengths
$\Gamma_{\al\beta}^\mu=2\pi N_{\mu}t_{\al\mu}t_{\beta\mu}^*$, where $N_\mu$
is the density of states in lead $\mu$. The AB flux enters via the phases of these complex tunneling matrix
elements, $t_{\al\mu}$, such that
$t_{1L}^*t_{1R}t_{2R}^*t_{2L}= t_{1L}t_{1R}^*t_{2R}t_{2L}^* e^{i2\varphi}$
where  $\varphi=2\pi\Phi/\Phi_0$,  $\Phi$ is the magnetic flux threading
the interferometer, and $\Phi_0=hc/e$ is the flux quantum.
The last term describes the electrostatic interaction between QD2 and the QPC, manifested in additional scattering potential when an electron occupied QD2.

\begin{figure}
  \includegraphics[width=3.5in]{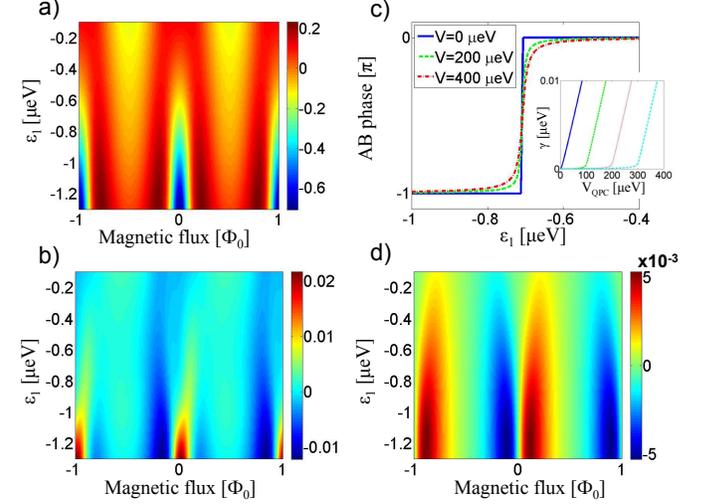}\\
  \caption{(color online) a) Linear response conductance as a function of magnetic flux (horizontal axis) and the energy of the level in the reference arm, $\eps_1$(vertical axis) for $V_{QPC}=0 \mu eV$, and b) its change when the QPC bias is $V_{QPC}=400 \mu eV$. (Other parameters are $\Gamma_{11}^L=\Gamma_{22}^L=1 \mu eV$,
$\Gamma_{11}^R=\Gamma_{22}^R=5 \mu eV$, $g_{c}=1.32\times 10^{-4}$, $\eps_2=1.5\mu eV$, $\eps_F=0 \mu eV$.) c) Phase of AB oscillations as a function of $\eps_1$ for different values of QPC bias; Since $\eps_2>\eps_F$, the transmission coefficient is not symmetric in respect to the Fermi level and the phase jump is shifted to $\eps_1=-0.7\mu eV$; Inset: dephasing rate, $\gamma(\eps_F)$ as a function of the QPC bias, $V_{QPC}$; different traces correspond to (from left to right) $\eps_2=1.5 \mu eV, 100 \mu eV, 200\mu eV$, and $300 \mu eV$;   d) Odd part of the AB conductance at $V_{QPC}=400 \mu eV$.}\label{fig:colorplot}
\end{figure}

The current via the interferometer was calculated using standard
techniques.\cite{MeirWingreen,Aguado} Interaction with the QPC was treated as
a second order self-energy  in QDs Green's functions.
The linear response
conductance  for zero bias on the QPC is shown in Fig.~\ref{fig:colorplot}a.
The AB oscillations are even in magnetic flux, and there is a phase jump that occurs around $\eps_1\approx -0.7\mu eV$,
where the oscillations change from having a minimum at zero magnetic field
to having a maximum.

When a finite bias is applied to the QPC, the phase symmetry is broken. Fig.~\ref{fig:colorplot}b depicts the difference in the conductance between  $V_{QPC}=400 \mu eV$ and $V_{QPC}=0$. The asymmetric component of the AB oscillations is now evident.
The amplitude of the odd component of the oscillations, shown in
Fig.~\ref{fig:colorplot}d, now takes experimentally measurable values.
The phase of the main harmonic of the AB oscillations, extracted by Fourier transform, is shown in Fig.~\ref{fig:colorplot}c for different values of $V_{QPC}$. This phase
changes abruptly at zero bias, but flows smoothly between $0$ and $\pi$ as the bias on the QPC increases.

The magnitude of the odd component of the AB oscillations is proportional to the strength of the coupling between the QPC and QD2, which also determines the experimentally observed reduction of the visibility of the AB oscillations in "which path?" experiments.\cite{Buks} Therefore it should also be possible to observe breaking of the phase symmetry experimentally.
Then the antisymmetric component of the AB oscillation  which can be extracted by anti-symmetrizing the data,\cite{PullerMeirPR} can be used to deduce the transmission phase through QD1,  $\varphi_{QD1}(\eps)$,  as a function of its energy, as discussed below.

Assuming that the level in QD2 is far from the Fermi level of the interferometer leads, i.e. $\Gamma_{11},\Gamma_{22},|\eps_1-\eps_F|<<|\eps_2-\eps_F|$,
one can express the odd component of the AB  conductance as
\begin{widetext}
\bq G^{odd}(\eps_F,\varphi)\simeq \pm sin\varphi \frac{G_0}{2}\frac{\gamma(\eps_F)\sqrt{\Gamma_{22}^L \Gamma_{22}^R}\left|\Gamma_{22}^L-\Gamma_{22}^R\right|} {2\left(\Gamma_{22}+\gamma (\eps_2)\right) \left[\left(\eps_2-\eps_F\right)^2+\frac{1}{4}\left(\Gamma_{22}+\gamma (\eps_2)\right)^2\right]}
\Re\left[t_{QD1}(\eps_F)-t_{QD1}(\eps_2)\right],\label{eq:Godd}\eq
\end{widetext}
where
$\gamma(\eps)$ is the additional broadening (dephasing rate) of the level in QD2 due to the  coupling to the QPC,\cite{Aleiner,gamma} and $\Re$ stands for the real part. Under the conditions outlined above. the visibility of AB oscillations is unaffected by the dephasing, and  $\gamma(\eps_2)$ in the denominator can be ignored   along with $\Gamma_{22}$, with respect to $(\eps_2-\eps_F)^2$.
In the numerator $\gamma(\eps_F)$ determines the  strength of the odd component of the AB oscillations.  At zero bias on the QPC $\gamma(\eps_F)$ is zero, which reflects the fact that no breaking of the phase symmetry occurs in an interferometer coupled to an equilibrium environment.\cite{SanchezKang} With increasing the QPC bias $\gamma(\eps_F)$ grows slowly till it reaches the "ionization threshold," $V_{QPC}=|\eps_2-\eps_F|$, after which it grows nearly linearly with bias (inset in Fig. \ref{fig:colorplot}c).

Similarly, the term proportional to $\Re\left[t_{QD1}(\eps_2)\right]$ in Eq.~(\ref{eq:Godd}) can be omitted along with $\gamma(\eps_2)$.
Note also that $G^{odd}$ vanishes in a symmetric device, i.e. when $\Gamma_{22}^L=\Gamma_{22}^R$. Other mechanisms of phase symmetry breaking are also known for their sensitivity to the device asymmetry.\cite{Buttiker2,PullerMeirPR,Leturcq,Device_asymmetry}

The only factor in Eq.~(\ref{eq:Godd}), dependent on the energy of the level in the reference arm of the interferometer, $\eps_1$, is the real part of the transmission amplitude $t_{QD1}(\eps_F)$, which is proportional to $\cos\varphi_{QD1}$, the sought after transmission phase. The other unknown energy-dependent quantities in this equation can be eliminated by measuring of the conductance via QD1 with QD2 disconnected,
\bq G_{QD1}(\eps_F)=\frac{G_0}{2} \left|t_{QD1}(\eps_F)\right|^2.\label{eq:GQD1}\eq
Then the cosine of the transmission phase via QD1 can be immediately extracted as
\bq \cos{\left[\varphi_{QD1}(\eps_F)\right]}\propto \frac{G^{odd}}{\sqrt{G_{QD1}}}.\label{eq:cos}\eq

Eq.~(\ref{eq:Godd}) is also correct for a multilevel QD in the reference arm, given that the abovementioned conditions  are still satisfied. This is true even in presence of interactions, as long as the interaction induced level broadening is small compared to $\Gamma_{11},\Gamma_{22}$.

\begin{figure*}[tbp]
  \includegraphics[width=7in]{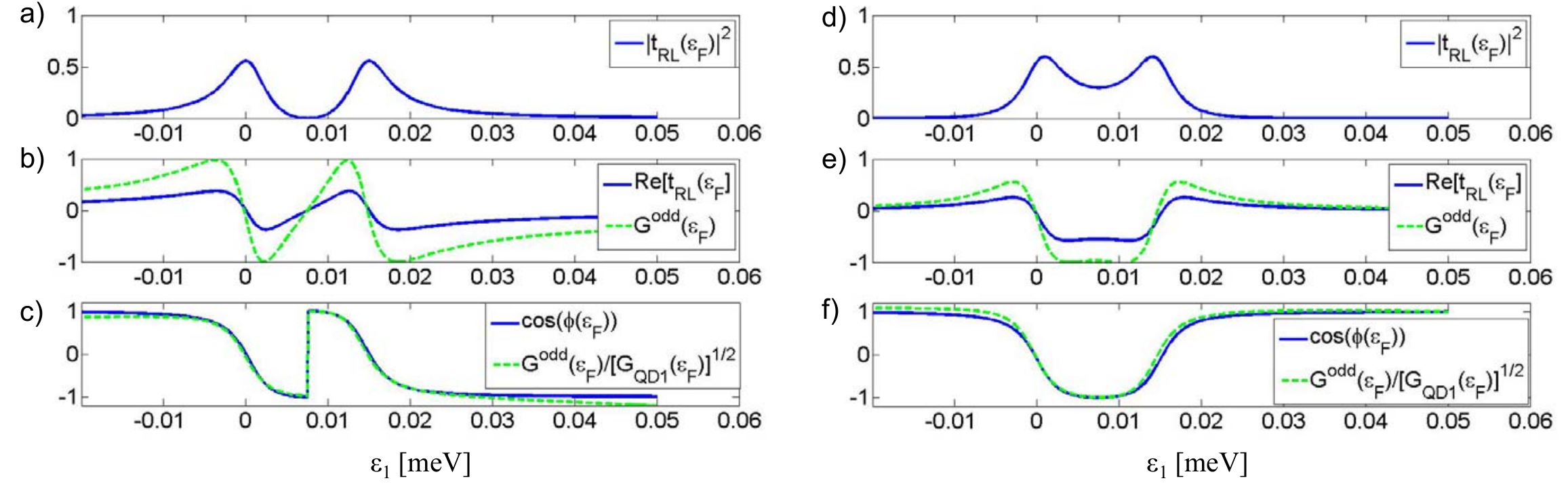}\\
  \caption{(color online) (a,d) Transmission, (b,e) real part of the transmission coefficient and odd AB conductance (normalized to its maximum), and (c,f) cosine of the transmission phase and its value extracted from $G^{odd}$ (Eq.~(\ref{eq:Godd})) for QD1 with two levels of the same parity (left) or different parity (right) (see parameters in the text).}\label{fig:sameparity}
\end{figure*}

Fig.~\ref{fig:sameparity} depicts the case of two levels of energies $\eps_{1a}$ and $\eps_{1b}=\eps_1-15\mu eV$ and identical widths $\Gamma_{11}^L=1 \mu eV$, $\Gamma_{11}^R=5 \mu eV$, having the same  (left panel) or opposite (right panel) parity. ($\eps_2=0.2 meV$, all other parameters are the same as those used for Fig.~\ref{fig:colorplot}. Level is defined to have even/odd parity, if the tunneling matrix elements connecting the level to the two leads, $t_{\al L}$ and $t_{\al R}$, have the same/opposite signs.) We see that the amplitude of the odd AB conductance (shown normalized to its maximal value) differs from the real part of the transmission coefficient only by a constant factor (and possibly sign), Fig.~\ref{fig:sameparity}b,e, as expected
from Eq.~(\ref{eq:Godd}). There is an excellent agreement between the transmission phase deduces from Eq.~(\ref{eq:cos}) and the one calculated directly, for both cases (Fig.3c and 3f).

The case of two levels of the same parity, shown in the left panel of Fig.~\ref{fig:sameparity}, is of particular interest in connection to the studies of phase lapses\cite{PhaseLapses} (apart from the fact that we do not account here for the Coulomb interaction) due to its important generic feature: an abrupt change of the phase by $\pi$, as can be seen in Fig.~\ref{fig:sameparity}c.

The abrupt change in the phase may occur only when the transmission coefficient, Fig.~\ref{fig:sameparity}a, is identically zero, i.e. when both its real and imaginary components vanish simultaneously, and then the phase is undefined. On the other hand, if only the real part is zero, the phase is $\pi/2$ and usually corresponds to a transmission resonance.
The fact that the zeros of $G^{odd}$ correspond to one of these two types of special points is important, since in principle, the transmission zeros of QD1 may be shifted in respect to those of QD1 with QD2 uncoupled, which may lead to unphysical results when naively applying Eq.~(\ref{eq:cos}). Therefore, one may have to shift $G^{odd}$ along the energy axis in order to make its zeros coincide with those of $G_{QD1}$, which we did when plotting Fig.~\ref{fig:sameparity}b,c.

In addition, if the interactions in QD1 are important, e.g., when studying Kondo effect,\cite{JI,Zaffalon,PhaseKondo} it may be impossible to characterize QD1 by its complex transmission amplitude and express its conductance by Eq.~\ref{eq:GQD1}, but the notion of the phase shift still remains relevant, and its special points can be identified from the zeros of $G^{odd}$, which are classified depending on whether $G_{QD1}$ takes zero value simultaneously with $G^{odd}$ or not.

In the case of two levels of different parity, Fig.~\ref{fig:sameparity}, right panel, the transmission is never equal to zero,  and the phase changes smoothly.

Since Eq.~(\ref{eq:cos}) is only a proportionality relation, the cosine of the transmission phase has to be normalized to interval $[-1,1]$. In the case of the levels of the same parity it is conveniently done using the abrupt change of the phase between $0$ and $\pi$, where the value of cosine should be set to change between $1$ and $-1$. In absence of such an abrupt jump, e.g. for the two levels of different parities, the cosine may be normalized by its peak value, Fig.~\ref{fig:sameparity}f, where phase is expected to take value $\pi$.

The normalization factor is the pre-factor in Eq.~(\ref{eq:Godd}), which is proportional to the dephasing rate and therefore dependent on the QPC bias, $V_{QPC}$. On the other hand, the transmission coefficient through QD1 is independent on this bias, i.e. by measuring $G^{odd}$ at different values of $V_{QPC}$ one should obtain results that differ only by a factor. Thus, having determined $\Re[t_{QD1}]$ in one measurement, one can use measurements at different values of $V_{QPC}$ to study the dependence of the dephasing rate, $\gamma(\eps_F)$, on $V_{QPC}$. If, in addition, one is able to measure independently the coupling strengths $\Gamma_{\al\al}^\mu$, one may use Eq.~(\ref{eq:Godd}) to extract the interaction constant $g_c$ and and compare it with that measured by other methods.\cite{Coupling}

To conclude,
we propose that breaking of the phase symmetry, necessary for measuring the transmission phase via a QD, embedded in an arm of an AB interferometer, can be achieved by coupling the interferometer to a QPC in a "Which path?" geometry, which is equivalent to phase symmetry breaking  by coupling to a non-equilibrium environment, predicted in Ref.~\onlinecite{SanchezKang}.
Although the phase of the resulting AB oscillations is not the transmission phase via the QD, the latter can be extracted from the amplitude of the odd part of the AB oscillations.

The authors would like to thank Klaus Ensslin and Thomas Ihn for useful discussions.
This work was supported in part by the ISF and BSF.

\end{document}